\documentclass[10pt,twocolumn]{article}

\usepackage[T1]{fontenc}
\usepackage{lmodern}
\usepackage[
  letterpaper,
  top=0.75in,
  bottom=0.85in,
  left=0.75in,
  right=0.75in
]{geometry}
\usepackage[hyphens]{url}
\usepackage{graphicx}
\usepackage{natbib}
\usepackage{caption}
\usepackage{amsmath}
\usepackage{amssymb}
\usepackage{algorithm}
\usepackage{algorithmic}
\usepackage{booktabs}
\usepackage[section]{placeins}
\usepackage[hidelinks]{hyperref}

\title{SkillSieve: A Hierarchical Triage Framework for Detecting Malicious AI Agent Skills}
\author{
    Yinghan Hou\\
    Department of Electrical and Electronic Engineering\\
    Imperial College London, London, United Kingdom\\
    \texttt{yh24@ic.ac.uk}
    \and
    Zongyou Yang\\
    Dyson School of Design Engineering\\
    Imperial College London, London, United Kingdom\\
    \texttt{zy2926@ic.ac.uk}
}
\date{}

\begin{document}
\maketitle

\begin{abstract}
Agent skills combine natural-language instructions with executable code while inheriting an agent's filesystem, credential, and network access. Attacks can span prose and files, whereas regex and code-only analyzers cover only one modality. \textsc{SkillSieve} applies three progressively deeper layers: recall-oriented regex, AST, and metadata triage; four parallel LLM security sub-tasks; and an independent three-model jury with debate on disagreement.

We evaluate 49,592 real ClawHub skills, a 390-skill labeled benchmark, and 100 adversarial samples across five evasion techniques on a \$440 ARM board. The full pipeline achieves F$_1=0.929$ (precision 0.912, recall 0.945) at an average \$0.006 per skill. An optional XGBoost fast-path reduces Layer-2/3 calls by 32\% for 1.7 F$_1$ points without changing recall. On 52 Feishu/Lark packages, Layer~2 removes 13 of 14 Layer~1 flags as domain-specific false positives; we also deploy the system as a Feishu chat bot. Code, labels, and aggregate results are open-sourced.
\end{abstract}

\section{Introduction}
\label{sec:intro}

AI coding agents such as OpenClaw~\cite{openclaw2026} extend their behavior through \emph{skills}: natural-language \texttt{SKILL.md} instructions plus optional scripts. Public registries such as ClawHub~\cite{clawhub2026} use automated screening and post-publication moderation rather than mandatory human pre-publication review, yet installed skills inherit agent privileges. A malicious package can therefore combine prompt injection with credential, filesystem, or network access~\cite{instructionhierarchy2024,injecagent2024,zdayagents2024}.

Detection must cover both prose and code. Static scanners miss instructions and cross-file intent; a single LLM provides no independent uncertainty signal. Following cost-aware LLM cascades~\cite{frugalgpt2023}, \textsc{SkillSieve} filters obvious cases cheaply, decomposes suspicious cases into four semantic tasks, and sends only high-risk cases to a debating multi-model jury.

Our contributions:
\begin{itemize}
    \item A three-layer static--semantic--jury pipeline that allocates analysis by risk (\S\ref{sec:framework}).
    \item Structured Semantic Decomposition into four parallel, independently inspectable tasks (\S\ref{sec:layer2}).
    \item A three-model jury with debate and conservative human-review routing (\S\ref{sec:layer3}).
    \item A 49,592-skill archive, 390-skill reviewed benchmark, and five-family adversarial evaluation on \$440 edge hardware (\S\ref{sec:dataset}, \S\ref{sec:eval}).
    \item Evaluation on 52 Feishu/Lark packages and deployment as a Feishu chat bot (\S\ref{sec:cross-ecosystem}).
\end{itemize}

\section{Background and Related Work}
\label{sec:background}

\subsection{AI Agent Skill Ecosystems}

The canonical skill contains YAML metadata and natural-language instructions in \texttt{SKILL.md}, plus optional Python, Bash, or JavaScript under \texttt{scripts/}~\cite{clawhub2026format}. ClawHub is a low-barrier GitHub-backed registry~\cite{clawhub2026,hkcert2026}; our snapshot contains 49{,}592 skills across roughly 16{,}800 publisher namespaces. Conventional package ecosystems already face LLM-induced package-hallucination risks~\cite{packagehallucinations2024}; skills additionally combine code with instructions executed under agent privileges, including credential, filesystem, and network access~\cite{ohm2020backstabber,instructionhierarchy2024,injecagent2024,agentworkflowthreats2025}.

\subsection{Attack Landscape}

Audits report prompt injection, typosquatting, cross-file credential exfiltration, and infostealers in utility skills~\cite{trendmicro2026}. ClawHavoc accounted for 335 of 341 malicious findings in one 2,857-skill audit~\cite{clawhavoc2026}; other studies report 13.4\% critical issues or 26.1\% overall vulnerabilities on different corpora~\cite{snyk2026toxicskills,liu2026agentskillswild}. These attacks cross the code/prose boundary.

\subsection{Existing Detection Approaches}

ClawVet combines static patterns with metadata, dependency, and typosquatting checks~\cite{clawvet2026}, while SkillFortify adds formal code analysis~\cite{skillfortify2026}. VirusTotal Code Insight and SkillProbe add LLM or multi-agent analysis~\cite{virustotal2026,skillprobe2026}, while SkillScan offers an optional neural classifier~\cite{skillscan2026}, amid broader work on indirect prompt injection, tool hijacking, skill attacks, and cyber-capable agents~\cite{toolhijacker2025,cybench2024,skillclone2026,skillject2026,skilltester2026}. Surveys cover the wider threat taxonomy~\cite{llmcybersurvey2025,agentworkflowthreats2025}. \textsc{SkillSieve} instead combines risk-based depth, four structured semantic tasks, and cross-model debate.

\section{Threat Model}
\label{sec:threat}

\textbf{Attacker.} An adversary publishes a skill to steal credentials, exfiltrate data, execute remote code, or socially engineer the agent. Evasion includes encoding, cross-file splitting, conditional or delayed triggers, and homoglyph substitution.

\textbf{Defender.} Before installation, the defender can inspect all text files and call remote LLMs but never executes the package. The objective is high recall with manageable false positives and API cost~\cite{vanrijsbergen1979,elkan2001}.

\textbf{Scope.} We detect malicious intent in text-only packages. Runtime monitoring, dynamic analysis, binaries, and payloads visible only after execution are out of scope.

\section{The SkillSieve Framework}
\label{sec:framework}

\subsection{Overview}

\begin{figure}[t]
\centering
\includegraphics[width=\linewidth]{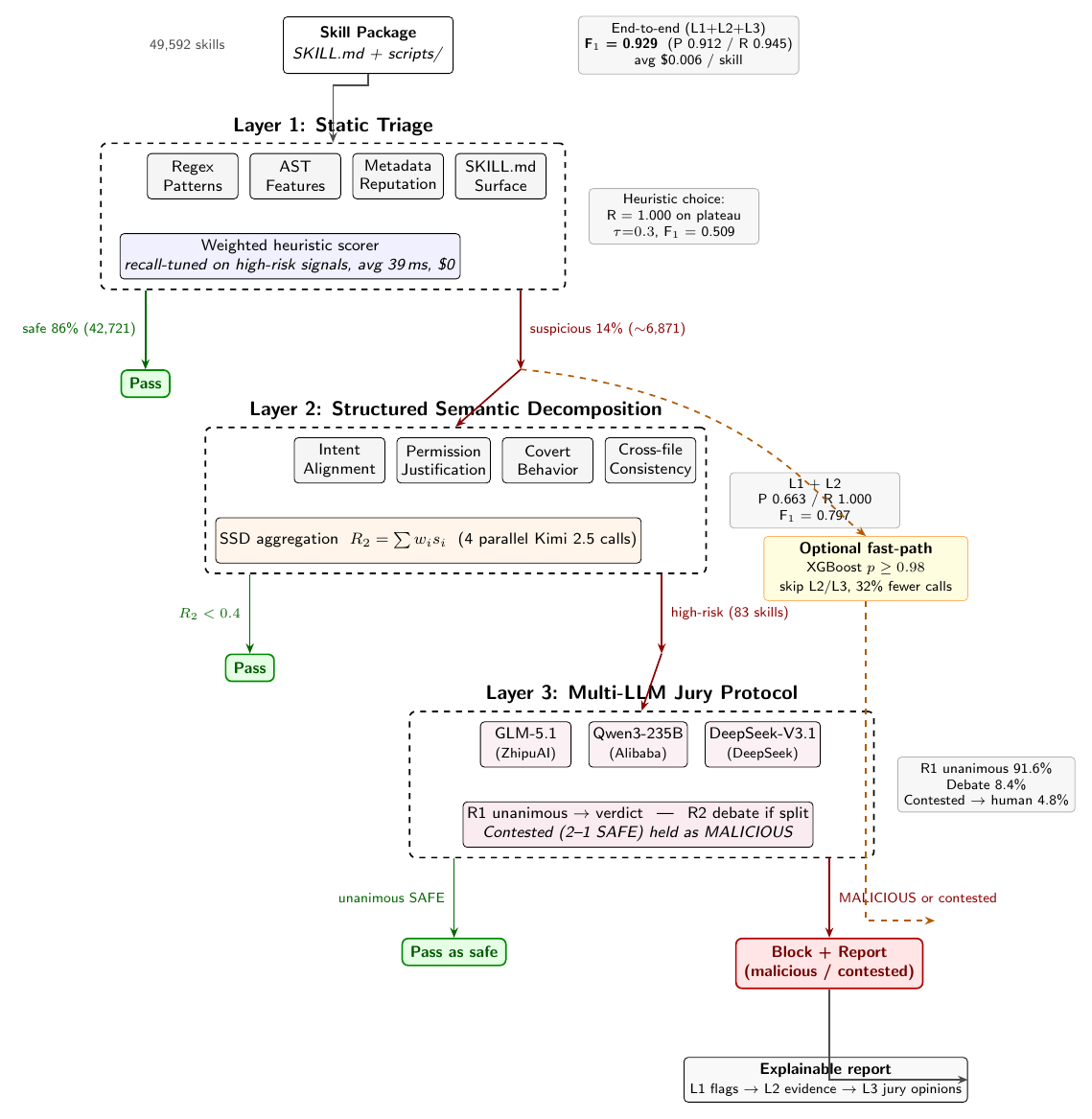}
\caption{The \textsc{SkillSieve} three-layer triage architecture. Layer~1 filters $\sim$86\% of skills via static analysis at zero cost. Layer~2 applies four parallel LLM sub-tasks to suspicious skills. Layer~3 convenes a multi-LLM jury for high-risk cases.}
\label{fig:architecture}
\end{figure}

\textsc{SkillSieve} applies progressively deeper analysis (Figure~\ref{fig:architecture}): Layer~1 combines four static modules with a recall-tuned heuristic; Layer~2 evaluates four semantic dimensions in parallel; and Layer~3 uses independent multi-LLM voting followed, when needed, by debate. Layer~1 filters about 86\% of packages at zero API cost, reserving LLM analysis for suspicious cases.

Algorithm~\ref{alg:pipeline} shows the full pipeline, including the optional fast-path branch (\S\ref{sec:fastpath}).

\begin{algorithm}[t]
\caption{\textsc{SkillSieve} pipeline}
\label{alg:pipeline}
\begin{algorithmic}[1]
\REQUIRE skill $s$; thresholds $\tau_1, \tau_2, \tau_{\text{high}}$; jurors $J = \{j_1, j_2, j_3\}$
\ENSURE verdict $v \in \{\textsc{Safe}, \textsc{Malicious}, \textsc{Contested}\}$
\STATE $r \gets \textsc{StaticScore}(s)$ \COMMENT{Layer~1}
\IF{$r < \tau_1$} \RETURN \textsc{Safe} \ENDIF
\IF{fast-path enabled \textbf{and} $\textsc{XGBoost}(s) \geq \tau_{\text{high}}$}
    \RETURN \textsc{Malicious}
\ENDIF
\STATE $R_2 \gets \sum_{i \in \{A,B,C,D\}} w_i \cdot \textsc{SubTask}_i(s)$ \COMMENT{Layer~2 SSD, parallel}
\IF{$R_2 < \tau_2$} \RETURN \textsc{Safe} \ENDIF
\STATE $V_1 \gets \{j(s) : j \in J\}$ \COMMENT{Layer~3 round~1}
\IF{$V_1$ unanimous} \RETURN $V_1[0]$ \ENDIF
\STATE $V_2 \gets \textsc{DebateRevote}(J, V_1)$ \COMMENT{round~2}
\IF{$V_2$ unanimous} \RETURN $V_2[0]$ \ENDIF
\IF{$\geq 2$ jurors vote \textsc{Malicious} in $V_2$} \RETURN \textsc{Malicious} \ENDIF
\RETURN \textsc{Contested} \COMMENT{held as malicious, routed to human}
\end{algorithmic}
\end{algorithm}

\subsection{Layer 1: Static Triage}
\label{sec:layer1}

Layer~1 favors recall over precision so downstream layers can resolve its false positives.

\subsubsection{Module A: Pattern Matching}
About 60 case-insensitive YAML rules cover reverse shells, credential theft, exfiltration, obfuscation, and prompt injection.

\subsubsection{Module B: AST Feature Extraction}
We use Tree-sitter~\cite{treesitter} to parse Python, Bash, and JavaScript, then derive counts of system, network, environment, and dynamic-execution operations together with encoded-literal and string-entropy features. Parse failures still pass through Module~A rather than being discarded.

\subsubsection{Module C: Metadata Reputation}
From \texttt{SKILL.md} frontmatter, we measure name similarity to the 100 most popular skills and flag sensitive environment-variable requests and dangerous binaries.

\subsubsection{Module D: SKILL.md Surface Statistics}
Cheap, no LLM. Instruction length, external URL count, permission-request count, mentions of sensitive paths (\texttt{\~{}/.env}, \texttt{\~{}/.ssh}), urgency-language density (``immediately'', ``must'', ``do not tell''), and the ratio of instruction length to description length.

\subsubsection{Classification}
A weighted heuristic maps the static signals to $r \in [0,1]$. Skills with $r<\tau$ pass as safe; the rest reach Layer~2. We set $\tau=0.3$, the highest round-number threshold on the recall-1.0 plateau (the lowest malicious score is 0.35), because downstream layers cannot recover a skill discarded here.

XGBoost~\cite{chen2016xgboost} reaches $0.918\pm0.011$ F$_1$ on the hold-out but misses malicious skills at every tested threshold; GroupKFold by author drops it to $0.671\pm0.176$. We therefore retain the zero-miss heuristic at Layer~1 and use XGBoost only as the optional fast-path in \S\ref{sec:fastpath}.

\subsection{Layer 2: Structured Semantic Decomposition}
\label{sec:layer2}

\subsubsection{Motivation}
Static analysis misses prompt injection and social engineering in natural-language instructions, while a monolithic LLM verdict can overlook individual threat dimensions. We therefore decompose the judgment.

\subsubsection{Four Sub-Tasks}
Tasks A and C analyze instructions: \textbf{Intent Alignment} compares stated and instructed behavior, while \textbf{Covert Behavior} detects concealment and safety bypasses. Tasks B and D analyze capability and code:

\begin{itemize}
    \item \textbf{Permission Justification.} Are environment, file, network, and binary permissions necessary for the stated purpose?
    \item \textbf{Cross-File Consistency.} Do scripts and instructed commands match \texttt{SKILL.md}? This catches split logic and script-less external payloads.
\end{itemize}

All four run \emph{in parallel} via concurrent API calls, so total latency is the maximum single-task latency (typically 2--5 seconds), not the sum.

\subsubsection{Prompt Design}
Each prompt includes a security-analyst role, full skill content, Layer~1 flags, task-specific instructions, and a strict JSON schema for score, evidence, and rating.

\subsubsection{Aggregation}
Each sub-task returns a risk score $s_i \in [0, 1]$. The Layer~2 risk score is a weighted sum:
\begin{equation}
    R_2 = w_A \cdot s_A + w_B \cdot s_B + w_C \cdot s_C + w_D \cdot s_D
\end{equation}
where $(w_A,w_B,w_C,w_D)=(0.35,0.25,0.25,0.15)$; skills with $R_2\geq0.4$ reach Layer~3.

\subsection{Layer 3: Multi-LLM Jury Protocol}
\label{sec:layer3}

\subsubsection{Motivation}
LLM judges exhibit systematic biases~\cite{mtbench2023}, and recent auditing shows that changing the judge can change the measurement itself~\cite{judgechange2026}. Multiple model families therefore expose disagreement while preserving an auditable cross-model check.

\subsubsection{Two-Round Protocol}

\textbf{Round 1: Independent Voting.} GLM-5.1, Qwen3-235B, and DeepSeek-V3.1 independently inspect the skill and prior-layer evidence, returning structured verdicts. Unanimity is final.

\textbf{Round 2: Structured Debate.} On disagreement, jurors review one another's evidence and revote~\cite{multiagentdebate2023}. Unanimity or a 2-of-3 \texttt{MALICIOUS} vote is accepted; a 2-of-3 \texttt{SAFE} split is held as malicious and sent to human review, reflecting the higher cost of false negatives.

\subsubsection{Explainable Reports}
Malicious verdicts include attack type, a three-layer evidence chain, and a recommended action.

\section{Dataset Construction}
\label{sec:dataset}

\subsection{Data Sources}

We construct our evaluation dataset from four sources:

\begin{itemize}
    \item \textbf{ClawHub full archive}: We clone the \texttt{openclaw/skills} GitHub repository (April 4, 2026 snapshot), which archives all skills published on ClawHub. This yields 49,592 skill packages across approximately 16,800 publisher namespaces.\footnote{We count the first path component in the archive as a publisher namespace. This is a repository-level identity count, not a count of verified individual people; one person or organization may control multiple namespaces.}
    \item \textbf{Snyk ToxicSkills}: The \texttt{snyk-labs/toxicskills-goof} repository provides documented malicious skill samples with known attack payloads~\cite{snyk2026toxicskills}.
    \item \textbf{ClawHavoc samples}: Malicious skills from the ClawHavoc campaign, identified via the \texttt{prompt-security/clawsec} security advisory feed and cross-referenced with the ClawHub archive.
    \item \textbf{Human-reviewed set}: 390 evaluated skills (55 malicious, 335 benign). All 390 labels received human review across multiple rounds (2026-04-14, 2026-04-17, 2026-04-28), followed by a final source-level audit on 2026-07-16. That audit corrected one high-risk but non-malicious developer-tool skill from malicious to benign.
\end{itemize}

\subsection{Labeling Schema}

Each skill receives three annotations: (1)~a binary label (\texttt{benign}/\texttt{malicious}); (2)~attack type multi-labels from a taxonomy of seven categories (prompt injection, credential theft, remote execution, data exfiltration, typosquatting, obfuscation, social engineering); and (3)~a stealth rating from 1 (plaintext malicious commands) to 5 (advanced obfuscation with conditional triggers).

\subsection{Adversarial Test Set}

We construct adversarial samples covering five bypass techniques: encoding obfuscation, cross-file logic splitting, conditional triggers, homoglyph substitution, and time-delayed payloads. Each sample combines its evasion technique with a credential theft payload injected into a benign skill template. Table~\ref{tab:adversarial} analyzes the per-layer interception pattern for each technique. We verify these patterns at scale with 100 samples (20 per technique) in Section~\ref{sec:eval}.

\section{Evaluation}
\label{sec:eval}

\subsection{Experimental Setup}

\textbf{Environment.} All experiments run on an Orange Pi AIpro single-board computer (4-core ARM64 CPU, 24\,GB LPDDR4X RAM, Ubuntu 22.04, Python 3.11). This hardware was chosen deliberately: it costs \$440 and represents the low end of what a developer might have on hand. Layer~1 analysis runs entirely on-device. Layers~2 and~3 call LLM APIs over WiFi: Kimi~2.5 (Moonshot AI) for Layer~2, and a three-family jury for Layer~3 (GLM-5.1 from ZhipuAI, Qwen3-235B from Alibaba, and DeepSeek-V3.1 from DeepSeek) chosen so no jury model overlaps with Layer~2. The evaluation dataset is the full \texttt{openclaw/skills} GitHub archive cloned on 2026-04-04.

\textbf{Baselines.} We compare against four baselines: (1)~ClawVet~\cite{clawvet2026} (v0.6.0, default non-semantic mode), combining 54 regex patterns with metadata, dependency, and typosquatting checks; (2)~SkillScan~\cite{skillscan2026} (v0.7.0, cached intelligence, optional neural ML disabled), with 138 static rules and 15 chain rules; (3)~SkillFortify~\cite{skillfortify2026}, a formal static analysis framework; (4)~VirusTotal Code Insight, a single-LLM (Gemini) analyzer.

\textbf{Metrics.} We report Precision, Recall, F$_1$, and False Positive Rate (FPR) for binary classification. Because our operating point treats missed threats as costlier than false alarms, we also report F$_2 = 5PR/(4P+R)$~\cite{vanrijsbergen1979,elkan2001}, which weights recall 2$\times$ as heavily as precision.

\subsection{Main Results}

\begin{table}[t]
\centering
\small
\setlength{\tabcolsep}{2pt}
\begin{tabular}{lccccc}
\toprule
\textbf{Method} & \textbf{P} & \textbf{R} & \textbf{F$_1$} & \textbf{F$_2$} & \textbf{FPR} \\
\midrule
ClawVet~\cite{clawvet2026} & 0.162 & 0.527 & 0.248 & 0.363 & 0.448 \\
SkillScan~\cite{skillscan2026} & 0.152 & \textbf{1.000} & 0.264 & 0.473 & 0.916 \\
SkillSieve L1 & 0.342 & \textbf{1.000} & 0.509 & 0.722 & 0.316 \\
\quad + SSD & 0.663 & \textbf{1.000} & 0.797 & 0.908 & 0.084 \\
\quad + SSD + Jury & \textbf{0.912} & 0.945 & \textbf{0.929} & \textbf{0.939} & \textbf{0.015} \\
\bottomrule
\end{tabular}
\normalsize
\caption{End-to-end detection on the 390-skill labeled benchmark (55 malicious, 335 benign). Metrics are recomputed from the archived per-sample predictions after the final label audit. The L1+L2 row has TP=55, FP=28, TN=307, FN=0. Layer~3 applies a contested-as-malicious policy (ties route to human review while holding the skill as malicious).}
\label{tab:main_results}
\end{table}

Layer~1 matches SkillScan's perfect recall with far fewer false positives. SSD then raises precision from 0.342 to 0.663 without reducing recall; the jury removes 23 of the remaining 28 false positives and routes four contested cases to human review.

\begin{figure}[t]
\centering
\includegraphics[width=\linewidth]{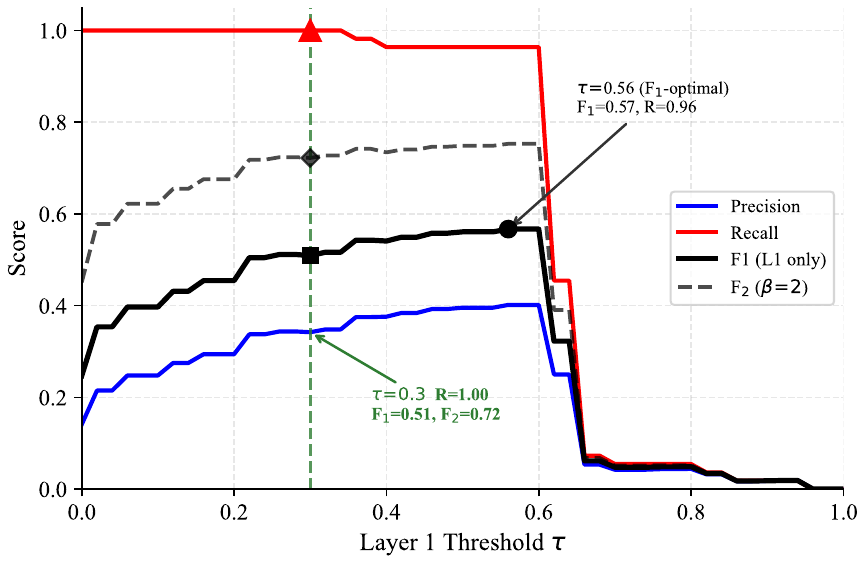}
\caption{Effect of Layer~1 threshold $\tau$ on precision, recall, F$_1$, and F$_2$ (L1 only). We choose $\tau=0.3$, the highest round-number threshold within the recall-1.0 plateau (the lowest malicious score in the benchmark is 0.35). At $\tau=0.3$, recall is 1.000, F$_1=0.509$, and F$_2=0.722$; raising $\tau$ past the plateau edge improves F$_1$ modestly but drops recall below 1.0, permanently losing malicious skills that Layer~2 cannot recover.}
\label{fig:threshold}
\end{figure}

\subsection{SSD Sub-Task Case Studies}

Three cases show complementary signals. \textbf{Cross-file Consistency} scores 0.95 on \texttt{auto-updater-\allowbreak nlt3m}, whose empty script directory conflicts with instructions to run external payloads. \textbf{Intent Alignment} scores 0.85 on \texttt{grok-scraper}, whose benign description masks Playwright anti-detection flags. \textbf{Permission Justification} and \textbf{Cross-File Consistency} both score 0.85 on \texttt{design-pick2}, whose undocumented script embeds credentials and exposes shell injection.

\subsection{Adversarial Robustness}

\begin{table}[t]
\centering
\footnotesize
\setlength{\tabcolsep}{2pt}
\begin{tabular}{p{1.6cm}ccp{2.5cm}}
\toprule
\textbf{Technique} & \textbf{L1 Score} & \textbf{Caught by} & \textbf{L1 Rule} \\
\midrule
Encoding & 0.35 & L1+L2 & obfuscation \\
Cross-file & 0.40 & L1+L2 & credential\_\allowbreak theft \\
Conditional & 0.70 & L1 & conditional\_\allowbreak trigger \\
Homoglyph & 0.80 & L1+L2 & prompt\_\allowbreak injection \\
Time-delay & 0.70 & L1 & time\_\allowbreak delay \\
\bottomrule
\end{tabular}
\normalsize
\caption{Per-layer interception analysis on five bypass techniques. Each technique combines its evasion method with a credential theft payload.}
\label{tab:adversarial}
\end{table}

Strong static signatures catch conditional and time-delayed payloads directly; semantic analysis confirms weaker encoding and cross-file signals, while metadata and content signals combine against homoglyph attacks. Across 100 generated variants (20 per technique), all 100 were detected.

\subsection{Efficiency Analysis}

\begin{table*}[t]
\centering
\begin{tabular}{lcccc}
\toprule
\textbf{Metric} & \textbf{CVet} & \textbf{SFort} & \textbf{VT} & \textbf{Ours} \\
\midrule
Avg local triage latency/skill & $\sim$1\,s & $\sim$5\,s & $\sim$3\,s & $\sim$40\,ms (Layer~1) \\
Estimated corpus-average API cost/skill & \$0 & \$0 & $\sim$\$0.01 & $\sim$\$0.006$^\dagger$ (full cascade) \\
GPU required & No & No & No & Optional \\
\bottomrule
\end{tabular}
\caption{Efficiency comparison. CVet = ClawVet, SFort = SkillFortify, VT = VirusTotal Code Insight.}
\label{tab:efficiency}
\end{table*}

\subsubsection{Cost Model}
$^\dagger$Let $c_i$ be each layer's per-skill cost, $p_2$ the probability of reaching Layer~2, and $p_3$ the conditional probability of reaching Layer~3. Then
\begin{equation}
\label{eq:cost}
\mathbb{E}[C] \;=\; c_1 \;+\; p_2\!\left(c_2 + p_3 \cdot c_3\right).
\end{equation}
With $c_1\approx\$0$, $p_2=0.14$, and $c_2+p_3c_3\approx\$0.04$, $\mathbb{E}[C]\approx\$0.006$. A fast-path resolving fraction $p_{\text{fp}}$ at cost $c_{\text{fp}}$ gives
\begin{equation}
\mathbb{E}[C_{\text{fp}}] \;=\; c_1 + p_2\!\left[(1{-}p_{\text{fp}})(c_2 + p_3 c_3) + p_{\text{fp}} c_{\text{fp}}\right],
\end{equation}
At $\tau_{\text{high}}=0.98$, this reduces downstream calls by 32\%. Estimated corpus cost falls from \$496 for single-LLM scanning to \$297.

\subsection{Edge Deployment Evaluation}

We evaluate Layer~1 on a \$440 Orange Pi AIpro without a GPU.

\begin{table}[t]
\centering
\begin{tabular}{lc}
\toprule
\textbf{Metric} & \textbf{Value} \\
\midrule
Total skills scanned & 49,592 \\
Total scan time & 1,863\,s (31.0\,min) \\
Avg latency / skill & 37.6\,ms \\
P95 latency / skill & 126.6\,ms \\
Skills flagged suspicious & 6,871 (13.86\%) \\
Errors (unparseable) & 1,623 (3.27\%) \\
Hardware cost & \$440 \\
API cost (Layer 1) & \$0 \\
\bottomrule
\end{tabular}
\caption{Layer~1 performance on Orange Pi AIpro (ARM64, 4-core, 24\,GB RAM) scanning 49,592 real ClawHub skills.}
\label{tab:edge}
\end{table}

Layer~1 processes all 49,592 skills in 31 minutes and sends only 6,871 to LLM analysis, a 7.2$\times$ reduction. This supports on-device triage in resource-constrained settings~\cite{edgellm2025,loraedge2026}.

\subsection{Jury Dynamics}

We analyze all 83 skills escalated by Layer~2 using the same jury and policy as Table~\ref{tab:main_results}.

\begin{table}[t]
\centering
\small
\setlength{\tabcolsep}{3pt}
\begin{tabular}{lcc}
\toprule
\textbf{Outcome} & \textbf{Count} & \textbf{Share} \\
\midrule
Unanimous Round~1 (no debate) & 76 & 91.6\% \\
Debate triggered (Round~2) & 7 & 8.4\% \\
\quad Unanimous after debate & 3 & \\
\quad Majority vote & 0 & \\
\quad Contested (escalated to human) & 4 & 4.8\% \\
\bottomrule
\end{tabular}
\normalsize
\caption{Layer~3 jury dynamics across all 83 L3-eligible skills in the main benchmark. Jury: GLM-5.1, Qwen3-235B, DeepSeek-V3.1.}
\label{tab:jury}
\end{table}

Seven cases require debate; three become unanimous and four remain 2--1 \texttt{SAFE} splits sent to human review. One contested case, \texttt{grok-scraper}, is malicious and would be missed by simple majority voting.

\subsection{Cost-Efficient Fast-Path Variant}
\label{sec:fastpath}

An optional XGBoost re-ranker accepts high-confidence malicious cases before Layer~2 to reduce LLM use.

\textbf{Method.} We train XGBoost~\cite{chen2016xgboost} on 1,438 skills after removing every benchmark path. Cases with $p\geq\tau_{\text{high}}$ skip Layers~2--3; all others follow the full pipeline.

\textbf{Calibration.} Five-fold GroupKFold by author fixes $\tau_{\text{high}}=0.98$ from out-of-fold predictions (precision 0.992; 129 TP, 1 FP), without consulting the benchmark. We disable a benign fast-path because every positive threshold loses recall when an author is held out.

\begin{table}[t]
\centering
\small
\setlength{\tabcolsep}{2.5pt}
\begin{tabular}{lccccc}
\toprule
\textbf{Variant} & \textbf{P} & \textbf{R} & \textbf{F$_1$} & \textbf{FPR} & \shortstack{\textbf{L2/L3}\\\textbf{calls}} \\
\midrule
Full pipeline & \textbf{0.912} & 0.945 & \textbf{0.929} & \textbf{0.015} & 161 \\
+ fast-path & 0.881 & 0.945 & 0.912 & 0.021 & \textbf{109} \\
\bottomrule
\end{tabular}
\normalsize
\caption{Full pipeline vs.\ cost-efficient fast-path variant ($\tau_{\text{high}} = 0.98$, OOF-calibrated) on the 390-skill benchmark. ``L2/L3 calls'' counts skills that reach Layers~2 or~3; fast-accepted skills skip both.}
\label{tab:fastpath}
\end{table}

The fast-path cuts downstream calls by 32\% at unchanged recall, while two additional false positives lower F$_1$ from 0.929 to 0.912; the full pipeline remains our headline result.

\textbf{Scope.} Because the benchmark is same-author-heavy and GroupKFold F$_1$ is 0.671, the 32\% saving is a same-distribution ceiling; new-author streams require recalibration and may save less.

\section{Cross-Ecosystem Generalization}
\label{sec:cross-ecosystem}

SkillSieve was built for OpenClaw, but the threat model is not OpenClaw-specific. We test generalization on the Feishu/Lark ecosystem, ByteDance's enterprise collaboration platform. Feishu has its own plugin and bot marketplace: the official CLI~\cite{larkcli2026} yields approximately 20 scannable packages identified by our adapter, the official OpenClaw channel plugin~\cite{openclawlark2026} bundles 9 more, and the community maintains dozens of connectors. The threat is real: CVE-2026-32974~\cite{cve202632974} allows unauthenticated attackers to inject forged Feishu events when webhook mode lacks an encryption key (CVSS 3.1 base score 9.8 per NVD), CVE-2026-25253~\cite{cve202625253} discloses authentication token exfiltration via a malicious \texttt{gatewayUrl} parameter, and \texttt{capability-evolver} was reported to contain undisclosed code capable of transmitting agent data to a hardcoded Feishu endpoint~\cite{capabilityevolver2026}. Enterprise platforms face the same supply chain risks as open registries.

\subsection{Adaptation}

We wrote a lightweight adapter that maps three Feishu package formats into the input expected by SkillSieve:
\begin{itemize}
    \item \textbf{\path{SKILL.md}-based skills} (\path{larksuite/cli}, \path{openclaw-lark}): parsed directly by the existing parser with zero modification.
    \item \textbf{OpenClaw plugins} (\texttt{openclaw.plugin.json} + source): metadata extracted from manifest and \texttt{package.json}; README.md used as skill description; TypeScript/JavaScript source files collected for pattern scanning.
    \item \textbf{Standalone bot repositories}: README.md mapped to skill description; all source files (Go, Python, JS/TS) collected as scripts.
\end{itemize}
We extended Layer~1 with 13 JavaScript/TypeScript patterns. They cover credential reads through \path{process.env}; outbound POST calls through \path{fetch} or \path{axios}; shell execution through \path{child_process.exec}; Feishu-specific endpoints such as \path{open.feishu.cn}; and hardcoded webhook URLs.

\subsection{Dataset}

We cloned 10 GitHub repositories spanning the Feishu ecosystem: \texttt{larksuite/cli} (official CLI, 20 skills), \texttt{larksuite/openclaw-lark} (official plugin, 9 skills), \texttt{m1heng/clawdbot-feishu} (community plugin, 9 skills), \texttt{yjwong/lark-cli} (alternative CLI, 8 skills), and 6 standalone bot/connector repositories. The adapter discovered \textbf{52 scannable packages} in total, a mix of official, community, and independent projects.

\subsection{Results}

Layer~1 classified 38 packages (73.1\%) as safe and flagged 14 (26.9\%) as suspicious, with an average scan time of 30.0~ms per package.

\textbf{True positives.} The \texttt{feishu-bridge} package stores \texttt{FEISHU\_APP\_SECRET} in a plaintext file at a predictable path (\texttt{\textasciitilde/.clawdbot/secrets/}), a genuine credential-storage concern surfaced by both layers.

\textbf{False positives from domain mismatch.} The remaining 13 flags came largely from three rule categories that interacted poorly with Feishu-specific content:
\begin{itemize}
    \item \emph{Time delay}: Unix timestamps in API documentation (e.g., \texttt{"end\_time": "1710086400"}) triggered the \texttt{time\_delay} rule, which was designed for Python's \texttt{time.time()} patterns.
    \item \emph{Obfuscation}: Long Feishu API URLs matched the base64-like regex designed to catch encoded payloads.
    \item \emph{Prompt injection}: \texttt{lark-mail}'s own anti-injection defense documentation contains the string ``Ignore previous instructions'' as an example of what to block.
\end{itemize}

\textbf{Layer~2 corrects Layer~1 false positives.} We ran Layer~2 (Kimi~2.5, SSD) on all 14 flagged packages. Layer~2 reclassified 13 as safe and kept only \texttt{feishu-bridge} flagged. For instance, \texttt{openclaw-lark/feishu-bitable} scored 0.350 at Layer~1 (suspicious) but only 0.065 at Layer~2 (safe): the LLM recognized the flagged string as a UI type enum in API documentation, not obfuscated code. The triage architecture is designed for this split of labor: Layer~1 over-flags on unfamiliar content; Layer~2 reads the context and filters out the noise.

\subsection{Deployment as a Feishu Chat Bot}

We deployed \textsc{SkillSieve} as a Feishu chat bot that enterprise users can talk to directly. The bot connects to the Feishu Open Platform via WebSocket long connection (no public IP required) and accepts two input modes: a GitHub URL (the bot clones, parses, and scans automatically) or pasted skill content. Results appear as a Feishu Interactive Card that updates live as the scan progresses: blue while scanning, then green (safe), orange (suspicious), or red (malicious) with full evidence.

Layer~1 finishes in under 1 second at zero cost for 73\% of Feishu packages. The remaining 27\% trigger Layer~2 LLM calls (cost $\sim$\$0.04/skill). Users can supply their own API keys; otherwise a default key handles up to 10 LLM scans per user per day.

\textbf{Security isolation.} The bot clones untrusted repositories, so we treat every input as hostile:
\begin{itemize}
    \item A URL whitelist accepts only HTTPS GitHub URLs, blocking local and non-GitHub schemes to prevent SSRF.
    \item Git hooks are disabled at clone time and the \texttt{.git/} directory is deleted immediately after checkout.
    \item A 100\,MB size cap rejects oversized repositories.
    \item All cloned content lives in an ephemeral temporary directory that is deleted after scanning.
    \item No code is ever executed; the scanner is read-only.
\end{itemize}

\textbf{Takeaway.} SKILL.md-based packages work with the existing parser unchanged. Non-standard formats need an adapter, but the core pipeline stays the same. Layer~1 flags more false positives on enterprise content (26.9\% vs.\ 14\% on OpenClaw) because its regex rules were tuned for Python/shell patterns, but Layer~2 reclassifies 13 of the 14 flagged packages as safe. The same adaptation applies to China's broader enterprise ecosystem. DingTalk (Alibaba, 700M+ users\footnote{DingTalk reached 700 million users by end of 2023 per Alibaba's official news outlet Alizila: \url{https://www.alizila.com/alibaba-dingtalk-700m-users-2023-ai-agent-boost-workspace-productivity/}.}), WeChat Work (Tencent's enterprise collaboration platform), QQ Bot, and Baidu Comate all expose plugin/bot frameworks with manifest-plus-code package structures. Adapting to each platform requires a format mapper and platform-specific pattern rules; the triage pipeline, SSD sub-tasks, and jury protocol carry over without modification.

\section{Discussion}
\label{sec:discussion}

\textbf{Limitations.} The main one is that Layer~1 only reads files. Payloads fetched from a remote URL at runtime are invisible to it, and time-delayed attacks are the hardest case across all methods because the malicious logic looks inert at scan time. Reported counts correspond to the archived evaluation artifacts; subsequent static-rule hardening in the released scanner may shift a small number of borderline Layer~1 false positives. Layers~2 and~3 inherit a different kind of uncertainty: LLM outputs are non-deterministic even at temperature 0, so the archived API run should be read as a fixed-run snapshot rather than a guarantee of bit-for-bit reproduction. There is also a more specific failure mode. Layer~2's structured decomposition tends to over-trigger on patterns that are inherent to a domain rather than malicious, and in our benchmark roughly a third of the Layer~2 false positives came from DeFi or wallet skills whose private-key handling and exchange-API calls read as suspicious one sub-task at a time. Domain-aware prompting and a max-based aggregation rule, which would escalate only when at least one sub-task produces a strong signal, look like the right directions to cut this class of false positive.

\textbf{Ethics.} Adversarial samples inject malicious logic into benign templates for evaluation only, and vulnerabilities we found in ClawHub during this work were reported to the OpenClaw security team before publication.

\textbf{What we would do next.} Runtime behavioral monitoring would catch the payloads static analysis misses. Fine-tuning a small open model on our labeled data could remove the API dependency for Layer~2. Cross-ecosystem transfer to Feishu/Lark is already working, and the remaining platforms (DingTalk, WeChat Work, QQ, MCP servers, LangChain tools) need only format adapters and platform-specific pattern rules. The area where more data would help most is Layer~1's cross-ecosystem false positive rate: scaling the labeled set past 390 skills and adding ecosystem-specific rules for JavaScript/TypeScript-heavy platforms should flatten it.

\section{Conclusion}
\label{sec:conclusion}

\textsc{SkillSieve} detects malicious agent skills by layering cheap static checks with focused LLM analysis and multi-model voting. On a 390-skill benchmark drawn from 49,592 real ClawHub skills, the pipeline achieves F$_1=0.929$ (precision 0.912, recall 0.945). Each layer plays a distinct role: Layer~1 filters 86\% of input at zero cost; Layer~2 SSD raises precision from 0.342 to 0.663 while retaining recall 1.000; the three-model jury (GLM-5.1, Qwen3-235B, DeepSeek-V3.1) acts as an independent quality check, raising precision to 0.912 and routing four contested cases to human review. Layer~1 scans all 49,592 skills on a \$440 ARM board in 31 minutes, while only 13.86\% are escalated to remote LLM analysis, yielding an estimated corpus-average API cost of \$0.006 per skill. All five tested adversarial bypass techniques (including conditional triggers, homoglyph typosquatting, and time-delayed payloads) are intercepted.

Beyond OpenClaw, a lightweight adapter scans 52 Feishu/Lark enterprise packages, where Layer~2 reclassifies 13 of Layer~1's 14 cross-ecosystem flags as safe. The same low-cost adaptation applies to DingTalk, WeChat Work, QQ, and other domestic enterprise platforms. We deploy \textsc{SkillSieve} as a production Feishu chat bot for real-time skill vetting, demonstrating that the triage architecture serves both offline batch scanning and interactive enterprise use.

\bibliographystyle{plainnat}
\bibliography{references}

\end{document}